\documentclass[aps,prb,twocolumn,floatfix,superscriptaddress]{revtex4-1}
\usepackage{graphicx,txfonts,bm}
\usepackage[colorlinks,citecolor=magenta,linkcolor=blue]{hyperref}

\begin{document}


\title{Emergence of quasiparticle multiplets in curium}


\author{Li Huang}
\email{lihuang.dmft@gmail.com}
\affiliation{Science and Technology on Surface Physics and Chemistry Laboratory, P.O. Box 9-35, Jiangyou 621908, China}

\author{Ruofan Chen}
\affiliation{Science and Technology on Surface Physics and Chemistry Laboratory, P.O. Box 9-35, Jiangyou 621908, China}

\author{Haiyan Lu}
\email{hyluphys@163.com}
\affiliation{Science and Technology on Surface Physics and Chemistry Laboratory, P.O. Box 9-35, Jiangyou 621908, China}

\date{\today}


\begin{abstract}
A combination of the density functional theory and the single-site dynamical mean-field theory is employed to study the electronic structures of various allotropes of elemental curium (Cm-I, Cm-II, and Cm-III). We find that the 5$f$ valence electrons in the high-symmetry Cm-I and Cm-II phases remain localized, while they turn into itinerancy in the low-symmetry monoclinic Cm-III phase. In addition, conspicuous quasiparticle multiplets are identified in the 5$f$ electronic density of states of the Cm-III phase. We believe that it is the many-body transition between $5f^{7}$ and $5f^{8}$ configurations that gives rise to these quasiparticle multiplets. Therefore, the Cm-III phase is probably a new realization of the so-called Racah metal.    
\end{abstract}


\maketitle


\section{Introduction\label{sec:intro}} 

The actinides, with atomic numbers ranging from 89 (actinium) to 103 (lawrencium) on the periodic table, have many complicated and fascinating properties. It is well accepted that, electronic structure regulates nonnuclear properties of materials. Since the actinide series successively fill up the 5$f$ shell, the role played by the 5$f$ electrons in the electronic structures of the actinides to the chemical and physical properties in their solid phases is at the heart of actinides science and is a subject of massive experimental and theoretical interest all the time~\cite{RevModPhys.81.235}. In spite of much efforts, there are still numerous open questions and puzzles, especially concerning with the entanglement between crystal structures and 5$f$ valence states of the actinides, that need to be answered and solved.  

Generally speaking, the actinides are often classified into two groups, early (or light) actinides and late (or heavy) actinides. In light actinides [from actinium (Ac) to plutonium (Pu)], the 5$f$ electrons tend to be itinerant and take actively part in chemical bonding, which leads to a gradual decrease in their atomic volumes. Though the 5$f$ electrons are capable of spin polarization and hence yielding some kinds of magnetic ordering states, magnetism is absent in most of light actinides~\cite{PhysRevB.72.054416}. As for heavy actinides [americium (Am) and beyond], the trend is exactly on the contrary. Their 5$f$ electrons incline to be localized, and there is no 5$f$ bonding. The localized 5$f$ electrons usually give rise to local magnetic moments. The sudden change in localization degree of freedom of 5$f$ electrons can explain the remarkable upturn in the atomic volumes of actinides~\cite{albers:2001}, i.e., the atomic volume of Am is almost 50\% larger than the one of its preceding neighbour Pu~\cite{LAReview}. In some sense, most of ground state properties of actinides could be understood or explained within this scenario. However, since the 5$f$ electronic states are incredibly sensitive to variation of external conditions and environment, the actinides will show quite intricate crystallographic phases and controversial solid properties under pressure, upon heating and alloying~\cite{RevModPhys.81.235}. For example, Pu, an element at odds with itself, comprises six allotropes which have different crystal structures and manifest distinct lattice properties~\cite{Hecker2004,LAReview,HECKER2004429}. Another interesting element is Am, which exhibits four crystal structures between ambient pressure and 100~GPa~\cite{PhysRevB.63.214101,PhysRevLett.85.2961,PhysRevB.84.075138}. 

Here, let us pay attention to curium (Cm), a pivotal element at the center of the actinide series. As a function of pressure, Cm will display five different allotropes and four successive phase transitions up to 100 GPa~\cite{Heathman110,HEATHMAN2007138}. At ambient pressure, the Cm-I phase is favorable. It crystallizes in a double hexagonal close packed structure. When pressure reaches 17 GPa, the Cm-I phase converts to the Cm-II phase. The latter is in a face centered cubic structure. The Cm-III phase has an atypical monoclinic structure with space group $C2/c$. The pressure range, that it is energetically favorable, spans from 37~GPa to 56~GPa. As pressure becomes larger, the Cm-IV phase appears. It has an orthorhombic structure with space group $Fddd$, which is similar to the Am-III phase~\cite{PhysRevB.63.214101}. Another orthorhombic structure with space group $Pnma$, the Cm-V phase, manifests itself above 95~GPa. It is analogous to the Am-IV phase~\cite{PhysRevB.63.214101}. Of particular interest with these allotropes and phase transitions are two folds. The first one is the occurrence of Cm-III. Its low-symmetry monoclinic structure is unique. It is absent in the high-pressure phases of the other heavy actinides, such as Am~\cite{PhysRevB.63.214101,PhysRevLett.85.2961}, Cf~\cite{PhysRevB.87.214111,PhysRevB.99.045109}, and Bk~\cite{HAIRE1984119}. Some people suggested that this lattice structure is stabilized by magnetism, or more specifically, the spin polarization of curium's 5$f$ electrons. Secondly, Cm's I$-$II and III$-$IV structural transitions are smooth, but its II$-$III and IV$-$V structural transitions are accompanied by significant volume collapses ($\sim$ 4.5\% and $\sim$ 11.7\%, respectively). Previous theoretical and experimental investigations suggested that these abrupt volume changes are due to the stepwise delocalization of Cm's 5$f$ electrons and their succeeding participation in chemical bonding~\cite{Heathman110,SODERLIND20142}.

Note that Cm is a highly radioactive and toxic element. It is not an easy task to carry out extensive experiments to study its electronic structures and the corresponding lattice properties~\cite{PhysRevB.99.224419}. Accordingly, theoretical calculations are necessary and become increasingly important in the last decades. Nowadays first-principles calculations based on density functional theory and its extensions could reproduce the experimentally observed sequence of phase transitions~\cite{Heathman110,SODERLIND20142}, if the 5$f$ electrons of Cm are assumed to be spin polarized and form antiferromagnetic long-range orders. It is predicted that Cm is a third element, besides iron and cobalt, in which energy associated with magnetic interaction influences the crystal structure of an element against pressure (or equivalently, volume)~\cite{Heathman110}. Latter this prediction is validated by experiments~\cite{HEATHMAN2007138}. Furthermore, the theoretical magnetic moment and X-ray absorption branching ratio for the cubic Cm-II phase agree quite well with the experimental values at ambient pressure~\cite{shim:2007}.

Though great progresses have been gained, it is worth pointing out that overall the electronic structures of Cm's five allotropes remain mysterious and almost untouched. Actually, to our knowledge, their band structures, Fermi surfaces, density of states, and 5$f$ valence states have not been studied systematically. We are not clear the similarities and differences in their electronic structures. We even do not understand why Cm's I$-$II and III$-$IV transitions are smooth. The underlying mechanism about, why the volume collapse in Cm's II$-$III transition is much smaller than that in Cm's IV$-$V transition, is also unknown. In order to provide reasonable explanations for the above questions, we employed the state-of-the-art first-principles many-body approach to study the electronic structures of Cm under moderate pressure. Our calculated results uncover that there will be a 5$f$ localized-itinerant crossover between the Cm-II and Cm-III phases. More important, we observe obvious signatures of quasiparticle multiplets in the 5$f$ electronic density of states in Cm-III. We further reveal that it is the valence state fluctuation and $5f^{7}-5f^{8}$ many-body transition that should be responsible for the emergence of quasiparticle multiplets in the Cm-III phase.

The rest of this manuscript is organized as follows. In Section~\ref{sec:method}, the first-principles calculations details are briefly introduced. Section~\ref{sec:results} is the major part of this manuscript. We present the calculated results and discussion in it. Section~\ref{sec:summary} serves as a short summary.   


\section{Method\label{sec:method}} 

In the present work, we utilized the density functional theory in combination with the single-site dynamical mean-field theory (dubbed as DFT + DMFT)~\cite{RevModPhys.78.865,RevModPhys.68.13} to study the electronic structures of Cm under pressure. The DFT + DMFT method is probably the most powerful first-principles approach that ever established for strongly correlated materials, and has been successfully applied to study the electronic structures of some actinides~\cite{PhysRevB.82.085117,PhysRevLett.101.126403,PhysRevB.75.235107,zhu:2013,PhysRevB.81.035105,Janoscheke:2015,dai:2003,PhysRevB.94.115148,shim_epl09,PhysRevLett.101.056403,savrasov:2001,PhysRevB.99.125113,rt:2019,joyce:2019,lh:2019}, including Cm metal in its cubic phase~\cite{shim:2007}. 

We used the \texttt{eDMFTF} package, which was implemented by K. Haule \emph{et al.}~\cite{PhysRevB.81.195107,wien2k}, to perform the charge fully self-consistent DFT + DMFT calculations. We used the experimental crystal structures of Cm under pressure~\cite{Heathman110}, and adopted the general gradient approximation (i.e., Perdew-Burke-Ernzerhof functional)~\cite{PhysRevLett.77.3865} to describe the exchange-correlation potential. The spin-orbit coupling effect was included. The system temperature was set to be 293~K, and the system was restricted to be paramagnetic. The 5$f$ orbitals of Cm atom were treated as correlated orbitals. The Coulomb interaction matrix was constructed by using the Slater integrals $F^{(k)}$. The Coulomb repulsion interaction parameter $U$ and Hund's exchange interaction parameter $J_{\text{H}}$ were 7.0~eV and 0.6~eV, respectively~\cite{RevModPhys.81.235}. The double-counting term for the self-energy function was represented by the fully localized limit scheme~\cite{jpcm:1997}. For the sake of simplicity, we ignored the non-equivalent Cm atoms in the Cm-I phase. In other words, all Cm atoms in Cm-I were assumed to be equivalent. The resulting multi-orbital quantum impurity models were solved by using the hybridization expansion continuous-time quantum Monte Carlo impurity solver (CT-HYB)~\cite{RevModPhys.83.349,PhysRevLett.97.076405}. We made a truncation in the local Hilbert space. Only those atomic eigenstates with $5f^{6}\sim 5f^{9}$ configurations were retained~\cite{PhysRevB.75.155113}. The Lazy trace evaluation trick was used to accelerate the calculations further~\cite{PhysRevB.90.075149}. The number of Monte Carlo sweeps was $2 \times 10^9$, which was enough to obtain converged results and suppress numerical noises~\cite{sign_problem}. Finally, the analytical continuations for self-energy functions were done by using the maximum entropy method~\cite{jarrell}.


\begin{figure}[th!]
\centering
\includegraphics[width=\columnwidth]{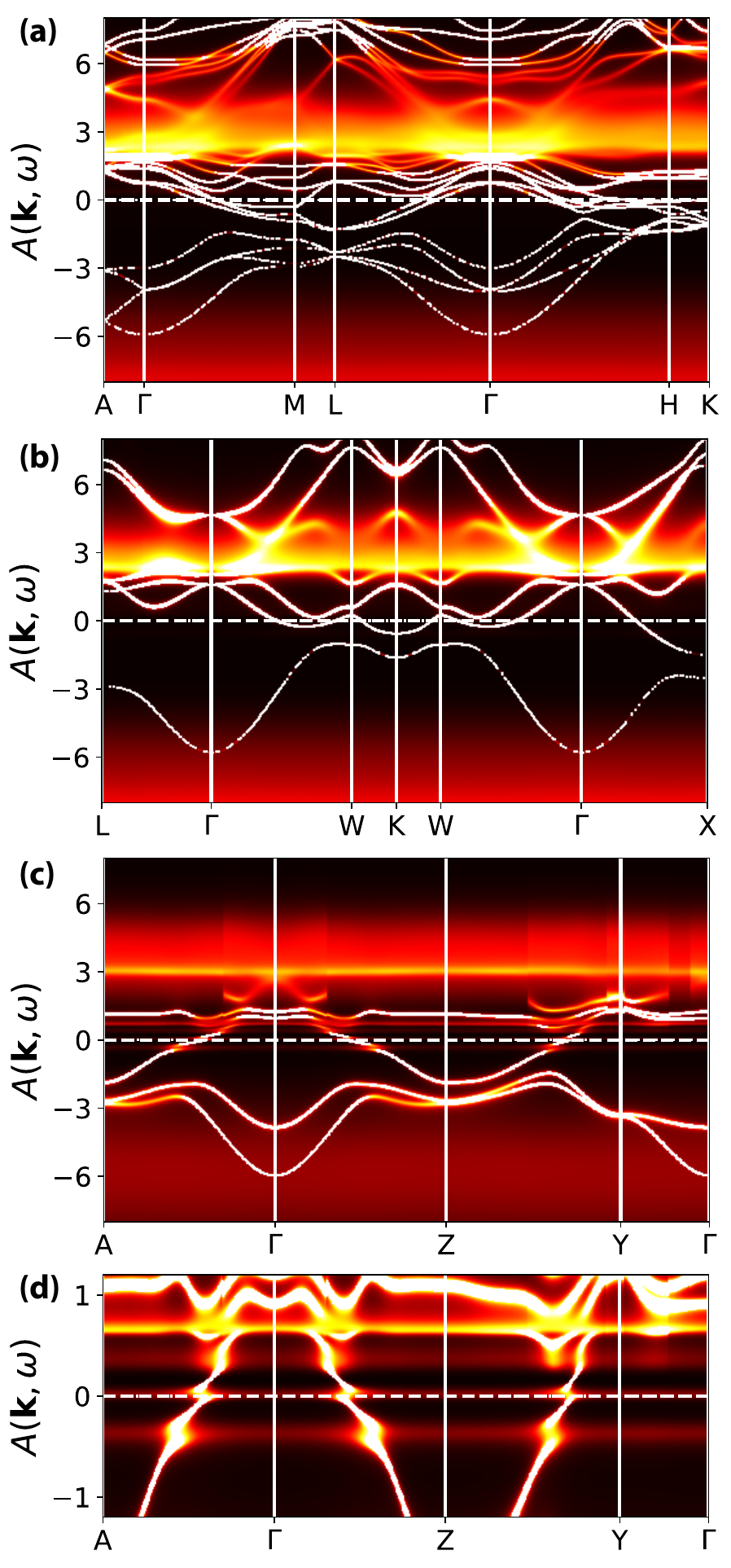}
\caption{(Color online). Quasiparticle band structures or momentum-resolved spectral functions $A(\mathbf{k},\omega)$ of Cm obtained by DFT + DMFT calculations. (a) Cm-I. (b) Cm-II. (c) Cm-III. (d) An enlarged view of panel (c) in the energy window $\omega \in$ [-1,1]~eV. In panels (d) and (c), the coordinates for the high-symmetry points are $A$~[0.0, 0.0, 0.5], $\Gamma$~[0.0, 0.0, 0.0], $Z$~[0.5, 0.0, 0.0], and $Y$~[0.0, 0.5, 0.0]. In these panels, the horizontal white dashed lines denote the Fermi levels. \label{fig:akw}}
\end{figure}

\begin{figure*}[ht]
\centering
\includegraphics[width=\textwidth]{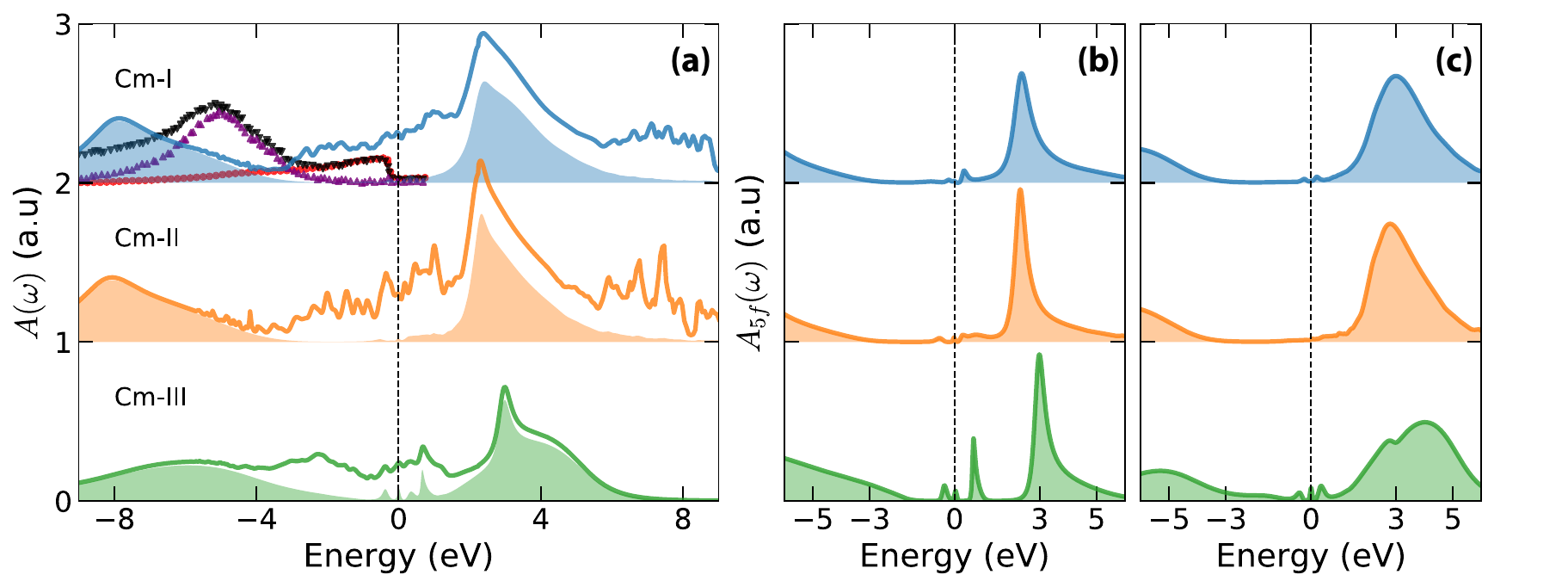}
\caption{(Color online). Density of states of Cm. (a) Total density of states $A(\omega)$ (represented by thick solid lines) and partial 5$f$ density of states $A_{5f}(\omega)$ (represented by colored-shadow areas). The experimental spectra for Cm-I are also shown in this figure. Here, filled red circles and black lower triangles denote the valence-band photoemission spectra measured with He-I and He-II radiation. The purple upper triangles denote the 5$f$ contribution to the Cm valence-band photoemission spectrum obtained by using He-II radiation. The original experimental data are extracted in Ref.~[\onlinecite{PhysRevB.83.125111}]. (b) and (c) The $5f_{5/2}$ and $5f_{7/2}$ components of partial 5$f$ density of states. Note that the data shown in these figures are rescaled for a better view. The vertical dashed lines denote the Fermi levels. \label{fig:dos}}
\end{figure*}

\begin{figure}[ht]
\centering
\includegraphics[width=\columnwidth]{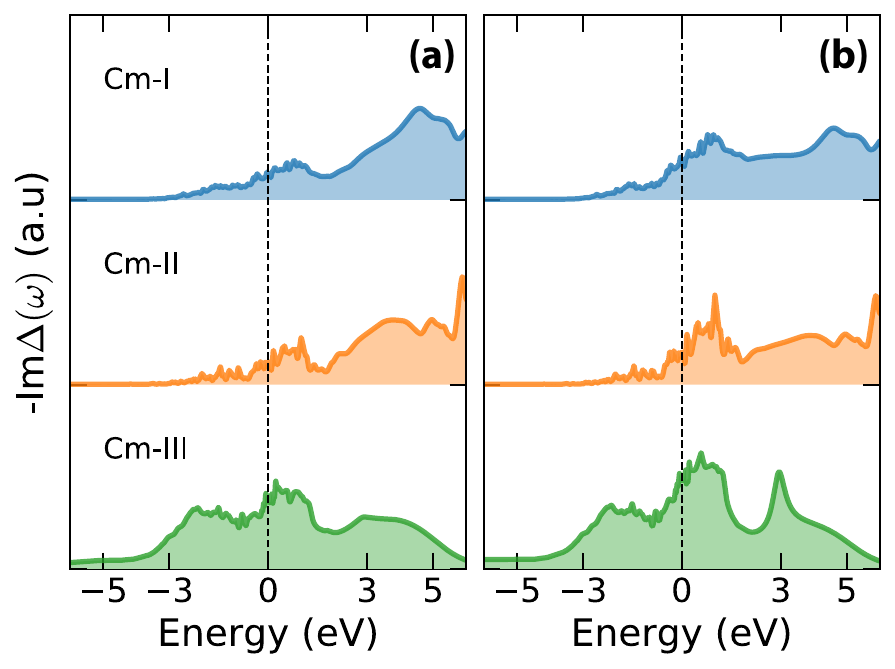}
\caption{(Color online). Imaginary parts of impurity hybridization functions of Cm's 5$f$ orbitals. (a) $5f_{5/2}$ component. (b) $5f_{7/2}$ component. The vertical dashed lines denote the Fermi levels. \label{fig:hyb}}
\end{figure}

\begin{figure*}[ht]
\centering
\includegraphics[width=\textwidth]{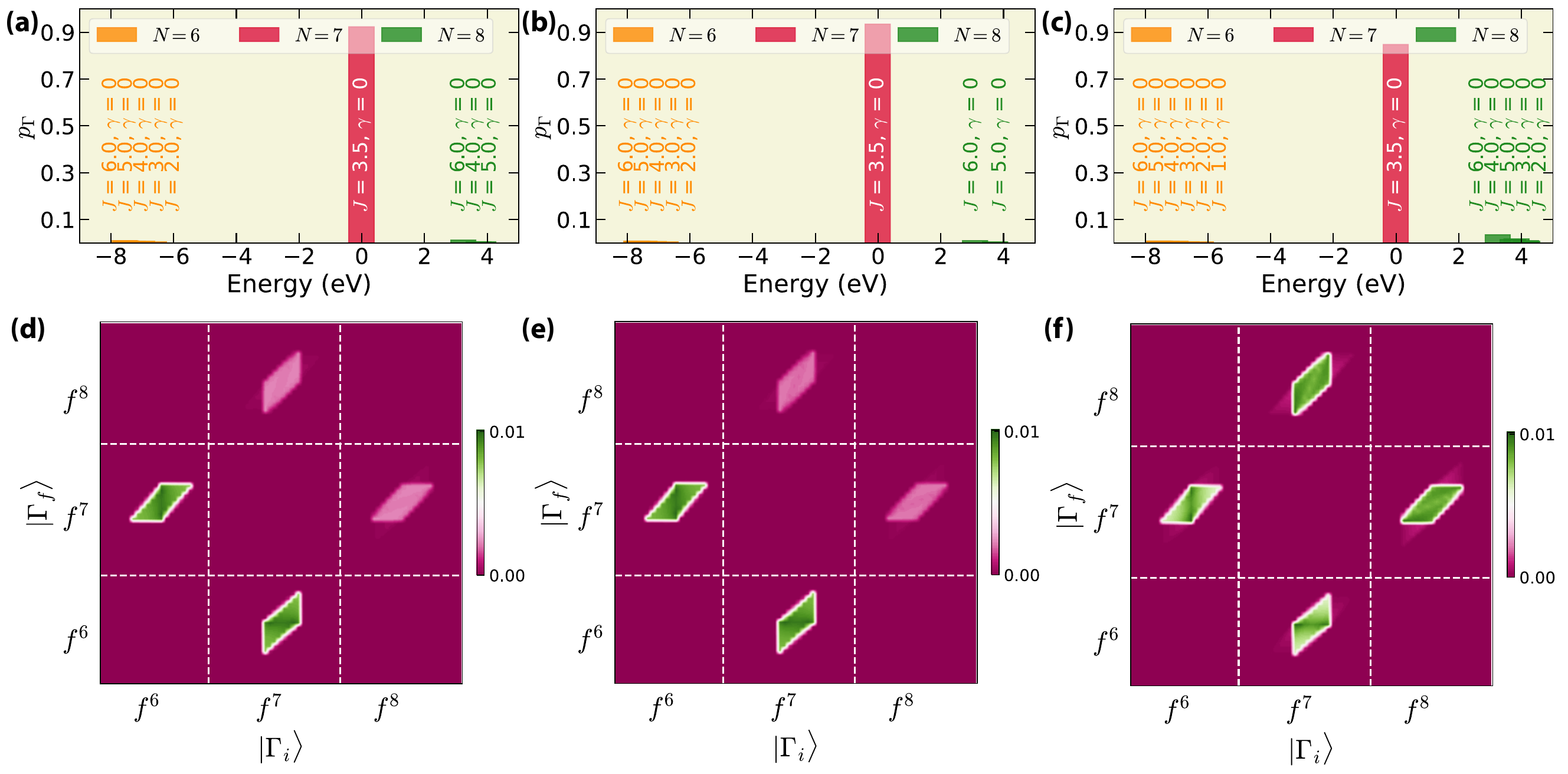}
\caption{(Color online). Valence state fluctuation in Cm. (a)-(c) Valence state histograms in the Cm-I, Cm-II, and Cm-III phases. Here, we used three good quantum numbers to label the atomic eigenstates. They are $N$ (total occupancy), $J$ (total angular momentum), and $\gamma$ ($\gamma$ stands for the rest of the atomic quantum numbers, such as $J_z$). Note that the contributions from the $5f^{6}$ atomic eigenstates are too trivial to be seen in these figures. (d)-(f) Transition probabilities between two different atomic eigenstates in the Cm-I, Cm-II, and Cm-III phases. Here $|\Gamma_i\rangle$ and $|\Gamma_f\rangle$ mean the initial and final atomic eigenstates, respectively. \label{fig:prob}}
\end{figure*}

\begin{table}[th!]
\caption{The weights for $5f$ electronic configurations $w(5f^{n})$, 5$f$ orbital occupancy ($n_{5/2}$, $n_{7/2}$, and $n_{5f}$, and X-ray absorption branching ratio $\mathcal{B}$ for the Cm-I, Cm-II, and Cm-III phases. See texts for more details. \label{tab:ratio}}
\begin{ruledtabular}
\begin{tabular}{rccccccc}
cases & $w(5f^{6})$ & $w(5f^{7})$ & $w(5f^{8})$ & $n_{5/2}$ & $n_{7/2}$ & $n_{5f}$ & $\mathcal{B}$ \\
\hline
Cm-I        & 4.31\% & 92.84\% & 2.85\% & 3.91 & 3.08 & 6.99 & 0.722\footnotemark[1] \\
            &        &         &        & 3.99 & 3.01 & 7.00 & 0.740\footnotemark[2] \\
            &        &         &        & 4.41 & 2.59 & 7.00 & 0.794\footnotemark[3] \\
            &        &         &        & 4.04 & 3.03 & 7.07 & 0.737\footnotemark[4] \\
            &        &         &        & 4.20 & 2.85 & 7.05 & 0.760\footnotemark[5] \\
\hline
Cm-II       & 3.81\% & 93.79\% & 2.40\% & 3.91 & 3.08 & 6.99 & 0.722\footnotemark[1] \\
            &        &         &        &      &      & 7.00 & 0.750\footnotemark[6] \\
            &        &         &        & 3.77 & 2.84 & 6.51 & 0.717\footnotemark[7] \\
\hline
Cm-III      & 4.27\% & 86.81\% & 8.91\% & 3.85 & 3.20 & 7.05 & 0.711\footnotemark[1] \\
\hline
Atomic LS   &        &         &        & 3.00 & 4.00 & 7.00 & 0.600\footnotemark[8] \\
Atomic IC   &        &         &        & 4.10 & 2.90 & 7.00 & 0.747\footnotemark[9] \\
Atomic $jj$ &        &         &        & 6.00 & 1.00 & 7.00 & 1.000\footnotemark[10]\\
\end{tabular}
\end{ruledtabular}
\footnotetext[1]{The present work. The 5$f$ impurity occupancy is calculated via the Matsubara Green's function $G(i\omega_n)$.}
\footnotetext[2]{See Ref.~[\onlinecite{PhysRevB.99.224419}]. Using the electron energy-loss spectroscopy and X-ray absorption spectroscopy.}
\footnotetext[3]{See Ref.~[\onlinecite{PhysRevB.76.073105}]. Using the electron energy-loss spectroscopy and X-ray absorption spectroscopy.}
\footnotetext[4]{See Ref.~[\onlinecite{PhysRevB.80.085106}]. Using the local density matrix approximation.}
\footnotetext[5]{See Ref.~[\onlinecite{PhysRevB.83.125111}]. Using the local density matrix approximation.}
\footnotetext[6]{See Ref.~[\onlinecite{shim:2007}]. Using the DFT + DMFT method.}
\footnotetext[7]{See Ref.~[\onlinecite{PhysRevB.94.115148}]. Using the DFT + DMFT method.}
\footnotetext[8]{See Ref.~[\onlinecite{RevModPhys.81.235}]. From many-electron atomic spectral calculations. The mechanism for angular momentum coupling is the LS scheme.}
\footnotetext[9]{See Ref.~[\onlinecite{RevModPhys.81.235}]. From many-electron atomic spectral calculations. The mechanism for angular momentum coupling is the IC scheme.}
\footnotetext[10]{See Ref.~[\onlinecite{RevModPhys.81.235}]. From many-electron atomic spectral calculations. The mechanism for angular momentum coupling is the $jj$ scheme.}
\end{table}

\section{results and discussion\label{sec:results}}

\subsection{Quasiparticle band structures} 

Let us concentrate on the quasiparticle band structures or momentum-resolved spectral functions $A(\mathbf{k},\omega)$ of Cm at first. In Fig.~\ref{fig:akw}, the momentum-dependent spectral functions of Cm along some selected high-symmetry lines in the first irreducible Brillouin zone are shown. 

For Cm-I and Cm-II [see Fig.~\ref{fig:akw}(a) and (b)], their quasiparticle band structures share some common characteristics. Firstly, the most striking features are the parallel and intensive stripe-like patterns around -6~eV and +3~eV. These stripes can be largely attributed to Cm's 5$f$ electrons. They resemble the upper and lower Hubbard bands of correlated 5$f$ electrons, respectively. Secondly, we observe noticeable band dispersions near the Fermi level. It means that they belong to the less-correlated $spd$ conduction electrons. The third, the 5$f$ electrons form huge band gaps (approximately 6~eV) in the two phases. And we hardly see any hybridization bands between the 5$f$ and $spd$ electrons in the vicinity of the Fermi level. So, it is concluded that the 5$f$ electrons in the Cm-I and Cm-II phases are completely localized and inert. 

As for Cm-III [see Fig.~\ref{fig:akw}(c)], the situation is a bit different. The stripe-like patterns still exist, but the original band gap between upper and lower Hubbard bands is greatly reduced (about 3 $\sim$ 4~eV). More important, we observe not only strong hybridizations between 5$f$ and $spd$ electrons near the Fermi level, but also a flat quasiparticle band which is exactly pinned at the Fermi level [see Fig.~\ref{fig:akw}(d)]. These features suggest that the 5$f$ electrons in the Cm-III phase are not completely localized any more, they become more and more itinerant and start to contribute to chemical bonding. Thereby, we anticipate that a 5$f$ localized-itinerant crossover~\cite{PhysRevB.99.045109} could occur when Cm goes from Cm-II to Cm-III. 

Finally, we would like to note that when pressure is increased (i.e., atomic volume is compressed) and temperature is decreased, the 5$f$ electrons tend to be itinerant (coherent). On the contrary, when pressure is reduced (i.e., atomic volume is expanded) and temperature is raised, the 5$f$ electrons lean toward the localized (incoherent) states. In an itinerant 5$f$ system, lattice distortion can easily split the narrow 5$f$ bands and thereby lower the total energy. Hence, low-symmetry structures are usually favored in the low-temperature phases of light actinides (such as the $\alpha$, $\beta$, and $\gamma$ phases of Pu~\cite{HECKER2004429,LAReview,Hecker2004}) and high-pressure phases of heavy actinides (such as the Cm-III, Cm-IV and Cm-V phases~\cite{HEATHMAN2007138,Heathman110}).

\subsection{Density of states} 

In Fig.~\ref{fig:dos}(a), we show the total density of states $A(\omega)$ and 5$f$ partial density of states $A_{5f}(\omega)$ of Cm. Though their crystal structures are very different, $A(\omega)$ and $A_{5f}(\omega)$ of the Cm-I and Cm-II phases are quite similar, just like what we have observed in their quasiparticle band structures. On one hand, their $A(\omega)$ always exhibit metallic characteristic. On the other hand, both $A_{5f}(\omega)$ show a large gap. However, $A_{5f}(\omega)$ of the Cm-III phase is surprising. We see that the itinerant-like Hubbard bands at high energy regime still exist, but the band gap disappears. There are several sharp and atomic-multiplets-like peaks near the Fermi level, instead of a single and fat quasiparticle resonance peak at the Fermi level. They are probably the quasiparticle multiplets, a concept firstly proposed by Yee \emph{et al.}~\cite{PhysRevB.81.035105}, who have found similar peaks in the 5$f$ electronic structures of plutonium chalcogenides and pnictides. The quasiparticle multiplets in these actinide compounds could be explained as a consequence of many-body transitions between the $5f^6$ and 5$f^5$ atomic multiplet configurations of Pu atom. In a previous work, we also identified quasiparticle multiplets in the low-temperature phases of metallic Pu~\cite{lh:2019}. Here, Cm-III is a new example that manifesting the feature of quasiparticle multiplets. Latter we will further discuss their underlying mechanism.

Due to spin-orbit splitting, the 5$f$ manifolds can be split into $5f_{5/2}$ and $5f_{7/2}$ components. In Fig.~\ref{fig:dos}(b) and (c), the $j$-resolved 5$f$ partial density of states are illustrated. Obviously, in the Cm-III phase, both $5f_{5/2}$ and $5f_{7/2}$ states contribute to the quasiparticle multiplets. However, for the Cm-I and Cm-II phases, there are almost featureless near the Fermi level. Another noticeable difference for these phases lies in the upper Hubbard bands. In Cm-III, the upper Hubbard bands are slightly shifted toward higher energy. Besides, the contribution from the $5f_{7/2}$ state is a double-peak structure, instead of a broad ``hump'' as is seen in Cm-I and Cm-II. We believe that this band splitting might originate from large lattice distortions in low-symmetry crystal structure of Cm-III~\cite{Heathman110,HEATHMAN2007138}. 

The electronic structure of Cm-I has been investigated by using photoemission spectroscopy a few years ago~\cite{PhysRevB.83.125111}. The experimental valence-band spectra are shown in Fig.~\ref{fig:dos}(a) as a comparison. The experimental spectra indicate full localization of the 5$f$ electrons, which are consistent with our theoretical results. We also discover sizable deviation between the theoretical and experimental peak positions for the lower Hubbard bands. We speculate that this divergence can be easily explained by the uncertainty of the Coulomb interaction parameters $U$ and $J_{\text{H}}$ used in the present DFT + DMFT calculations~\cite{PhysRevB.83.125111,RevModPhys.81.235}. Actually, if the values of $U$ and $J_{\text{H}}$ are rescaled by a factor, we can reproduce the experimental spectra.  

In Fig.~\ref{fig:hyb}, we show the imaginary parts of 5$f$ hybridization functions $-\text{Im} \Delta(\omega)$, which can be regarded as a measurement for the strength of $c-f$ hybridization~\cite{RevModPhys.78.865,RevModPhys.68.13}. From this figure, we can see that the hybridization between 5$f$ and $spd$ electrons in Cm-III is much larger than those in Cm-I and Cm-II around the Fermi level. It implies once again that the 5$f$ electrons in Cm-I and Cm-II are localized, but turn to be delocalized in Cm-III.  

\subsection{Valence state fluctuation}

Valence state fluctuation might be a common phenomenon for $f$-electron materials. It has been observed or predicted in many actinide-based materials or cerium-based heavy fermion materials~\cite{shim:2007,Janoscheke:2015,lh:2019}. Previous DFT + DMFT calculations regarding cubic phase Cm suggested that valence state fluctuation is very weak in Cm. Its 5$f$ occupancy is very close to the nominal value 7~\cite{shim:2007}. However, we think pressure may alter this picture, in other words, it is still possible to see strong valence state fluctuation and deviation from the nominal 5$f$ occupancy in the high-pressure phases of Cm. In order to examine this idea, we try to calculate the valence state histogram $p_{\Gamma}$ (i.e., atomic eigenstate probability) of Cm by using the CT-HYB quantum impurity solver~\cite{PhysRevB.75.155113,PhysRevB.81.195107}. Here, $\Gamma$ means the atomic eigenstates $| \Gamma \rangle \equiv |\psi_{\Gamma} \rangle$, which are labelled by using some good quantum numbers, such as total occupancy $N$ and total angular momentum $J$. Then $p_{\Gamma}$ denotes the probability to find out a valence electron in a given atomic eigenstate $|\Gamma\rangle$.

The calculated results for $p_{\Gamma}$ are presented in Fig.~\ref{fig:prob}(a)-(c). Just as expected, the valence state fluctuations in Cm-I and Cm-II are quite weak. The predominant atomic eigenstate is undoubtedly $|N = 7, J = 3.5, \gamma = 0 \rangle$. Its probability accounts for more than 90\%. The contributions from the other atomic eigenstates are too trivial to be seen in these figures. As for Cm-III, $|N = 7, J = 3.5, \gamma = 0 \rangle$ is still the principal atomic eigenstate. Though as a whole Cm-III is yet a system that exhibiting weak valence state fluctuation, the contributions from the other atomic eigenstates [mainly from $5f^{8}$ ($N = 8$) configurations] become quite important. We thus predict that the valence state fluctuation would become more and more important in Cm-IV and Cm-V, which usually stabilize under higher pressure~\cite{Heathman110}.

We also calculate the transition probability between arbitrary two atomic eigenstates, $\Pi(\Gamma_i | \Gamma_f)$, where $\Gamma_i$ and $\Gamma_f$ denote the initial and final states, respectively~\cite{PhysRevB.75.155113}. This observable can help us understand the nature of many-body transitions in Cm under pressure. The calculated results are shown in Fig.~\ref{fig:prob}(d)-(f). For Cm-I and Cm-II, the distributions of transition probabilities are fairly similar. The probabilities for $5f^{6}-5f^{7}$ transitions are much larger than those for $5f^{7}-5f^{8}$ transitions. This is quite natural because the weight of $5f^{6}$ configuration is larger than the one of $5f^{8}$ configuration in these phases. As for Cm-III, the transitions between $5f^{7}$ and $5f^{8}$ configurations become comparable to the transitions between $5f^{6}$ and $5f^{7}$ configurations, since the weight of $5f^{8}$ configuration is almost twice as large as the weight of $5f^{6}$ configuration (see Table~\ref{tab:ratio}). Thus, we suspect that the quasiparticle multiplets seen in the density of states of Cm-III likely originate from the many-body $5f^{7}-5f^{8}$ transitions.

\subsection{X-ray absorption branching ratio and 5$f$ occupancy}

How the 5$f$ electrons occupy the $5f_{5/2}$ and $5f_{7/2}$ levels across this series is a particularly fundamental question about actinides. In general, it is determined by the scheme of angular momentum coupling that each actinide exhibits. Depending on the relative strength of spin-orbit coupling and electrostatic interaction, the angular momenta of multi-electronic systems have three ways to couple with each other: Russell-Saunders (LS) coupling, $jj$ coupling, and intermediate coupling (IC)~\cite{RevModPhys.81.235}. For the ground states of late actinides, intermediate coupling is favorite. Previous theoretical and experimental researches already demonstrated that the Cm-I and Cm-II exhibit intermediate coupling. But, we immediately have a new question. How about the high-pressure phases of Cm? To this end, we try to evaluate the 5$f$ orbital occupancies for Cm-I, Cm-II, and Cm-III. The calculated values are summarized in Table~\ref{tab:ratio}, and are compared with previous experimental and theoretical results where available.  As for Cm-I, our data agree quite well with the very recent X-ray absorption and magnetic circular dichroism measurements~\cite{PhysRevB.99.224419}. We find that not only Cm-I, but also Cm-II, fulfill the requirement of intermediate coupling. Interestingly, the 5$f$ orbital occupancies for Cm-III show an obvious deviation from the intermediate coupling scheme. There is a slight shift toward the LS limit. We believe that such a deviation or shift is available in Cm-IV and Cm-V phases as well and could become more remarkable. Note that a similar trend has been suggested in the cubic phase Cf under pressure~\cite{PhysRevB.99.045109}.

With either electron energy-loss spectroscopy or X-ray absorption spectroscopy, a core electron is excited above the Fermi level, allowing to probe directly the unoccupied states. In Cm, due to strong spin-orbit coupling, the transitions from 4$d$ core states to 5$f$ valence states result in two absorption lines, i.e., $N_{5}$ ($4d_{5/2} \rightarrow 5f$) and $N_4$ ($4d_{3/2} \rightarrow 5f$). The relative strength of the $N_5$ absorption line is the so-called 
X-ray absorption branching ratio $\mathcal{B}$. It measures the strength of the spin-orbit coupling interaction in the 5$f$ shell. If we ignore the electrostatic interaction between 4$d$ and 5$f$ electrons, $\mathcal{B}$ can be evaluated via the following equation~\cite{shim:2007,shim_epl09}:       
\begin{equation}
\label{eq:ratio}
\mathcal{B} = \frac{3}{5} - \frac{4}{15} \frac{1}{14 - n_{5/2} - n_{7/2}} \left ( \frac{3}{2} n_{7/2} - 2 n_{5/2} \right ).
\end{equation}
Here, $n_{7/2}$ and $n_{5/2}$ represent the 5$f$ occupation numbers for the $5f_{7/2}$ and $5f_{5/2}$ states, respectively. The calculated results are shown in Table~\ref{tab:ratio} as well. We have $\mathcal{B}(\text{Cm-I}) = \mathcal{B}(\text{Cm-II}) > \mathcal{B}(\text{Cm-III})$. Since X-ray absorption branching ratio is sensitive to 5$f$ delocalization~\cite{PhysRevB.76.073105}, these results indicate one more time that the 5$f$ electrons in Cm-III is more delocalized than those in Cm-I and Cm-II. 


\section{Concluding remarks\label{sec:summary}} 

In the present work, we employed the DFT + DMFT method to study the electronic structures of Cm-I, Cm-II, and Cm-III. The major findings are as follows. At first, the 5$f$ electrons in Cm-I and Cm-II are completely localized with a large band gap. In Cm-III, rampant change occurs. The band gap is replaced by quasiparticle multiplets. Second, the 5$f$ electrons are virtually locked into the $5f^{7}$ configuration, especially in the Cm-I and Cm-II phases. There is only one overwhelming peak, which denotes the ground state of Cm atom ($|N = 7, J = 3.5, \gamma = 0 \rangle$), in the valence state histogram. As is compared to Pu, valence state fluctuation in Cm is rather weak. Third, many-body transitions between $5f^{7}$ and $5f^{8}$ become nontrivial in Cm-III. They boost the quasiparticle multiplets. Thus the 5$f$ spectra of Cm-III contain two distinct parts, well-pronounced atomic multiplet structures near the Fermi level and broad Hubbard bands at high energy regime. This is quintessential feature of ``Racah materials'' or ``Racah metals'', a concept proposed by A. B. Shick \emph{et al}~\cite{shick:2015,PhysRevB.87.020505}. Therefore, it is suggested that Cm-III is a material realization of the so-called ``Racah metal''. Finally, intermediate coupling scheme still approximately holds in Cm. But a trend toward the LS limit is observed in Cm-III. 

The present work can enrich our understanding about the 5$f$ electronic structures of actinides. According to the calculated results, we find that the electronic structures (including quasiparticle band structures, density of states, valence state histograms, and 5$f$ occupancy) of Cm-I and Cm-II are quite similar. It explains why Cm's I$-$II transition is so smooth. On the other hand, the electronic structures of Cm-III are somewhat different. Its 5$f$ electrons turn into (partially) delocalized, $c-f$ hybridization and valence state fluctuation are enhanced. These changes lead to the $\sim$ 4.5\% volume collapse during Cm's II$-$III transition. As for the large volume collapse observed in Cm's IV$-$V transition, it can be explained by a completely delocalization of 5$f$ electrons~\cite{Heathman110}. 

Finally, the use of DFT + DMFT method for studying electronic band structures and extracting valence state histograms of various allotropes of an element should have great applications for the other late actinides. Am, Cf, and Bk are several pressing examples. They exhibit complicated $V-T$ phase diagrams, which usually comprise multiple allotropes~\cite{PhysRevB.63.214101,PhysRevLett.85.2961,PhysRevB.87.214111,HAIRE1984119}. Most of lattice properties of these allotropes remain unclear. Further calculations are highly desired.

\begin{acknowledgments}
This work was supported by the Natural Science Foundation of China (No.~11874329, 11934020, and 11704347), the Foundation of President of China Academy of Engineering Physics (No.~YZ2015012), and the Science Challenge Project of China (No.~TZ2016004).
\end{acknowledgments}


\bibliography{cm}

\begin{thebibliography}{49}
\bibitem [{sig()}]{sign_problem}%
  \BibitemOpen
  \href@noop {} {}\bibinfo {note} {We meet severe negative sign problems when
  we try to apply the CT-HYB quantum impurity solver to solve the impurity
  models derived from the Cm-IV and Cm-V phases. So, in the present work, we
  have to restrict our studies for the Cm-I, Cm-II, and Cm-III phases
  only.}\BibitemShut {Stop}%
\end{thebibliography}

\end{document}